\documentclass{PoS}

\PoS{PoS(LAT2005)107}

\title{Improving the dynamical overlap algorithm }

\ShortTitle{Improving the dynamical overlap algorithm }

\author{Nigel Cundy\\
        Universi\"at Wuppertal\\
        E-mail: \email{cundy@theorie.physik.uni-wuppertal.de}}

\author{\speaker{Stefan Krieg}\\
        NIC/ZAM Forschungzentrum J\"ulich/Universi\"at Wuppertal\\
        E-mail: \email{s.krieg@fz-juelich.de}}

\author{Thomas Lippert\\
        NIC/ZAM Forschungzentrum J\"ulich/Universi\"at Wuppertal\\
        E-mail: \email{th.lippert@fz-juelich.de}}

\abstract{We present algorithmic improvements to the overlap Hybrid Monte Carlo algorithm, including preconditioning techniques and improvements to the correction step, used when one of the eigenvalues of the Kernel operator changes sign, which is now O($\Delta t^2$) exact.}

\FullConference{XXIIIrd International Symposium on Lattice Field Theory\\
         25-30 July 2005\\
         Trinity College, Dublin, Ireland}

\newcommand{\bra}[1]{\langle #1|}
\newcommand{\ket}[1]{|#1\rangle}

\newcommand{\Tr}[0]{\mathrm{Tr}}

\begin{document}




\section{Improved Correction Step\label{correctionstep}}
\begin{figure}
\begin{center}
\includegraphics[width=.8\textwidth]{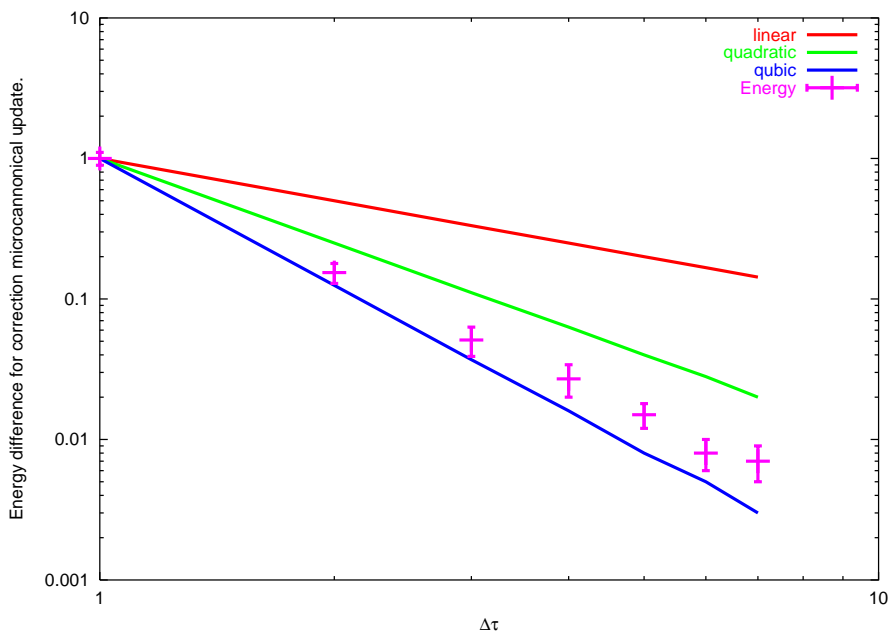}
\end{center}
\caption{Dependency of the energy on $\Delta\tau$. The red lines are from top down: ($\Delta\tau$, $\Delta\tau^2$, $\Delta\tau^3$).}
\label{energy}
\end{figure}

The sign function in the overlap Dirac operator creates a
discontinuity $-2\;d$ in the pseudo-fermion contribution to the action
whenever
an eigenvalue of the kernel operator changes sign. To conserve energy, we
integrate up to the computer time $\tau_c$ where the eigenvalue crosses, and
introduce a discontinuity in the kinetic energy which exactly cancels the
jump in the pseudo-fermion energy. A general area
conserving and reversible update which can do this is:
\begin{eqnarray}
\Pi^+&=&\Pi^-+\left( \eta
(\eta,\Pi^-)\right)\left(\sqrt{1+ \frac{4\;d_0}{
(\eta,\Pi^-)^2}}-1\right)+\sum_{j=1}^N
\left(\eta_1^j(\eta_1^j,\Pi^-)+\eta_2^j(\eta_2^j,\Pi^-)\right)
\nonumber\\
&&\times\left(\sqrt{1+
\frac{4\;d_j}{(\eta_1^j,\Pi^-)^2+(\eta_2^j,\Pi^-)^2}}-1\right) + A,\label{eq:1}
\end{eqnarray}
$\Pi_-$ is the original momentum, $\Pi_+$ the final momentum, A is an arbitrary function of the gauge field at $\tau_c$, $\eta$ is a
unit vector normal to the $\lambda=0$ surface, and the $\eta^i_j$ are
unit vectors normal to $\eta$. The original
algorithm \cite{Fodor} set $d_1=4\;d$ and $d_j=0$, and had
O($\tau_c$) errors. We can use the $d_j$ terms to cancel these errors,
giving the \textit{transmission} algorithm:
\begin{eqnarray}
\Pi^+ &=& \Pi^- + \tau_c(F) - \eta \tau_c(\eta,F) - \frac{\tau_c}{3}\Tr(F)+
(\eta,\Pi^-)\sqrt{1+\frac{4\;d}{(\eta,\Pi^-)^2}} \nonumber\\
&&+  \left(\eta^2_1 (\eta^2_1, \Pi^- - {\hat F}) + \eta_2^2 (\eta_2^2,
\Pi^-
- {\hat F})\right) \times\left(\sqrt{1 +
  \frac{d_2 }{(\eta^2_1,
    \Pi^- - {\hat F})^2 +  (\eta_2^2, \Pi^- -
    {\hat F})^2}}-1\right)
\nonumber\\
d_2 &=&  - 2\tau_c(F^-,\eta)(\Pi^-,\eta)
  +2\tau_c(F^+,\eta)(\Pi^+,\eta)+ 2\tau_c({\hat F},F^+ - F^-).\nonumber\\
{\hat F} &=& \frac{\tau_c}{2}(F^- + F^+) \nonumber
\end{eqnarray}
where $F^{\pm}$ are the MD forces immediately before and after the
crossing. We cannot use this algorithm if it would lead to complex $\Pi^+$. In this
case, we have to \textit{reflect} of the $\lambda=0$ surface, and there
will be no topological charge change.  
Figure \ref{energy} shows how the energy difference across the
correction step varies as a function of $\Delta\tau$. It clearly shows that the energy has errors of at maximum O($\Delta\tau^2$).

\section{Improved Leapfrog algorithm\label{leapfrog}}
In \cite{deForc} an alternative leapfrog update for the molecular dynamics part of the HMC is suggested:
\begin{enumerate}
\item $\Pi(\tau+\lambda\Delta\tau) = \Pi(\tau) + \lambda\Delta\tau \dot{\Pi}(\tau)$.
\item $U(\tau+\Delta\tau/2) = e^{i(\Delta\tau/2) \Pi(\tau+\lambda\Delta\tau)}U(\tau)$.
\item $\Pi(\tau+(1-\lambda)\Delta\tau) = \Pi(\tau) + (1-2\lambda)\Delta\tau \dot{\Pi}(\tau+\lambda\Delta\tau)$.
\item $U(\tau+\Delta\tau) = e^{i(\Delta\tau/2)   \Pi(\tau+(1-\lambda)\Delta\tau)}U(\tau+\Delta\tau/2)$.
\item $\Pi(\tau+\Delta\tau) = \Pi(\tau+(1-\lambda)\Delta\tau)+ \lambda\Delta\tau \dot{\Pi}(\tau+(1-\lambda)\Delta\tau)$.
\end{enumerate}
The optimal value of $\lambda$ is given in \cite{deForc}. This algorithm has improved energy conservation, which more than compensates for the need to invert the overlap operator twice. We have tested it on $4^4$, $8^4$, and $12^4$ lattices, and found gains of around 30$\%$ (see section \ref{results}).

\section{Stout Smearing\label{Smearing}}
We use the ``stout links'' proposed in \cite{stout}. As mentioned in \cite{deGrand} this improves the condition number of the Wilson operator substantially, thus speeding up the inversions needed to construct the overlap operator. We find however that there is a "phase transition" at a critical level of the smearing parameter, leading to a sharp increase in the magnitude of the smallest eigenvalue of the Wilson operator. This reduces the effectiveness of the smearing.

\section{Hasenbusch acceleration\label{hasenbusch}}
Hasenbusch acceleration has been used to speed up dynamical simulations. We introduce an additional fermion flavour with a large mass, and by placing the two fermions on different time scales we can in principle reduce the number of low mass inversions needed during a trajectory. However, we saw little gain when using this method, partly because we were testing on large masses, and partly because our overlap operators are usually well conditioned (see section \ref{results}).

\section{Overlap eigenmode preconditioning\label{preconditioning}}
\begin{figure}
\begin{center}
\includegraphics[width=.8\textwidth]{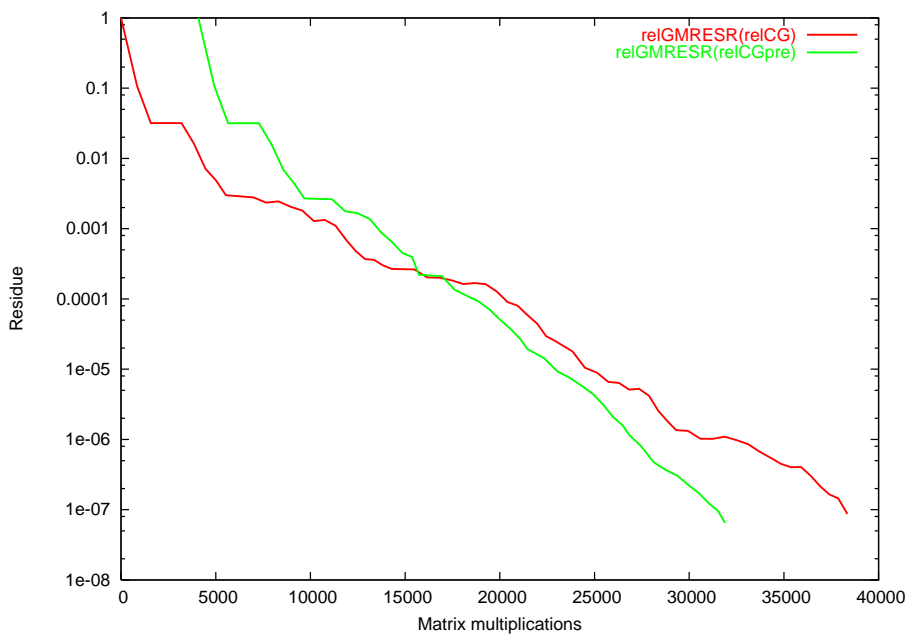}
\end{center}
\caption{Convergence history for the preconditioning method on configuration with trivial topology}
\label{plots_preconditioning_non-zero}
\end{figure}

\begin{figure}
\begin{center}
\includegraphics[width=.8\textwidth]{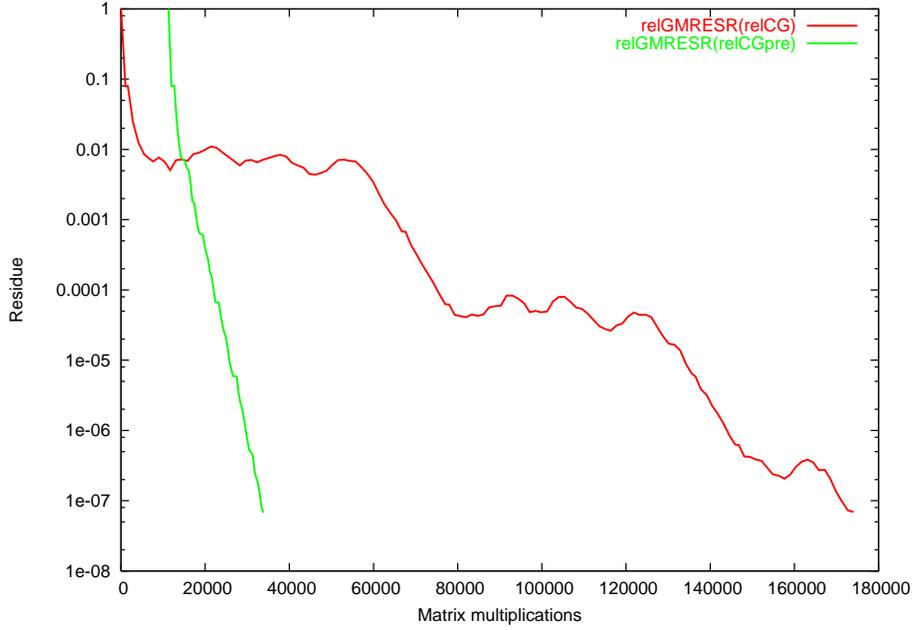}
\end{center}
\caption{Convergence history for the preconditioning method on configuration with non trivial topology}
\label{plots_preconditioning_zero}
\end{figure}

In the case of a topological nontrivial configuration, the spectrum of the overlap matrix includes a ``zero mode''. Inversions of the overlap operator become prohibitively expensive when simulating in the regime of small quark masses. Our ansatz is to calculate the smallest $m$ eigenvectors $\Psi_m$ and eigenvalues $\lambda_m$ of the overlap operator to a very low precision (e.g. $10^{-2}$) and use them as a preconditioner for our CG preconditioner in our GMRESR inverter. Our preconditioner is:
$$
P = 1+\sum_m\left(\frac{\alpha_m}{\lambda_m}-1\right)\ket\Psi_m\bra\Psi_m
$$
Figures \ref{plots_preconditioning_non-zero} and \ref{plots_preconditioning_non-zero} show the convergence of CG with and without preconditioning using above projector. These plots were generated using a $8^4$ dynamical configuration at mass $\mu=0.1$, with the inversions carried out at mass $\mu=0.03$. Figure \ref{plots_preconditioning_non-zero} shows the convergence history for the case of a configuration with trivial topology; Figure \ref{plots_preconditioning_zero} shows the convergence history for a configuration with a ``zero mode'' induced by topology: Clearly in the latter case the preconditioning offers great possible gains, which --- according to our experience --- increase with the volume and decreasing of the masses. 


In an HMC simulation, using the previous eigenvectors as a starting point for the next eigenvalue calculation can dramatically reduce the time needed, although it is unclear how large an effect this leads has on the reversibility of the MD.

\newpage
\section{Non area conserving correction step\label{nap}}
It is possible to use a non area conserving molecular dynamics update by including the Jacobian in the Metropolis accept/reject step\footnote{We thank A. Borici for pointing this out to us}. The detailed balance condition reads:
\begin{eqnarray}
P[U^\prime \leftarrow U]W_C[U]  &=& \int \mathrm{d} \Pi \mathrm{d} \Pi^\prime
\exp^{-\frac{1}{2}\Pi^2}\delta([U,\Pi]-T_{MD}[U^\prime,\Pi^\prime]) 
\min\left(1,\exp^{\Delta}\right) W_C[U]
\nonumber \\
& =& \int
\mathrm{d}\Pi\mathrm{d}\Pi^\prime\exp^{-\frac{1}{2}{\Pi}^2}\delta([U^\prime,-\Pi^\prime]-T_{MD}^{-1}[U,-\Pi])
\frac{\partial U^\prime, \Pi^\prime}{\partial U, \Pi}
\min\left(1,\exp^{\Delta}\right)W_C[U] \nonumber \\
& = &
\int\mathrm{d}\Pi\mathrm{d}\Pi^\prime\exp^{-\frac{1}{2}{\Pi^\prime}^2}\delta([U^\prime,-\Pi^\prime]-T_{MD}^{-1}[U,-\Pi])
\min\left(1,\exp^{-\Delta}\right)W_C[U^\prime] \nonumber\\
&=& P[U \leftarrow U^\prime]W_C[U^\prime]\nonumber \\
\Delta &=& -\ln \left[\frac{\partial U^\prime, \Pi^\prime}{\partial U,
\Pi}\right] +
S_G[U]+\frac{1}{2}\Pi^2-S_G[U^\prime]-\frac{1}{2}{\Pi^{\prime}}^2
\nonumber
\end{eqnarray}
The most general transmission update which is reversible and conserves $\Delta$ is:
\begin{eqnarray}
\exp^{-(\Pi^+,\eta)^2/2}      &=& \exp^{-(\Pi^-,\eta)^2-2d}-\exp^{-r_0^2/2 - 2d}+\exp^{-r_0^2/2} \nonumber
\end{eqnarray}
For $r_0=\infty$, this gives the usual area conserving transmission
formula equation (\ref{eq:1}). One has to reflect if the transmission formula gives a
complex ($\Pi^+,\eta$).
By tuning $r_0$, we can improve the transmission rate.
The results displayed in the tables of section \ref{results} were obtained using $r_0=1$, and give
a 50\% improvement in the transmission rate.




\section{Results\label{results}}
In this section we summarise the results referred to in the previous sections.

\begin{center}
\begin{tabular}{l|l|l|l|l|l|l}
Type& time& Acc & $n_{md}$& trans./traj.& refl./traj. &$n_{t} $ \\
\hline
normal   & 1897(60)& 94$\%$& 40 &  0.0738(240)& 1.348(100)& 325\\
has      & 1986(20)& 88$\%$& 40 &  0.0521(311)& 1.059(94) & 307\\
imp      & 1420(10)& 94$\%$& 15 &  0.0535(233)& 0.876(98) & 299\\
imphas   & 1594(40)& 75$\%$& 15 &  0.0772(336)& 1.093(118)& 324\\
impnap   & 1480(10)& 95$\%$& 15 &  0.117(34)  & 1.336(136)& 310\\
impnaphas& 1611(60)& 78$\%$& 15 &  0.110(21)  & 0.832(159)& 155\\
\hline
\end{tabular}
\\$\mu=0.05$
\end{center}

\begin{center}
\begin{tabular}{l|l|l|l|l|l|l}
Type& time& Acc & $n_{md}$& trans./traj.& refl./traj. & $n_{t} $\\ 
\hline
normal   & 1816(20)& 95$\%$& 40 &  0.447(64)& 0.938(80) & 465\\
has      & 2100(90)& 90$\%$& 40 &  0.569(65)& 0.880(65) & 374\\
imp      & 1479(20)& 96$\%$& 15 &  0.371(43)& 0.947(62) & 533\\
imphas   & 1470(60)& 90$\%$& 15 &  0.413(53)& 0.531(76) & 518\\
impnap   & 1445(50)& 95$\%$& 15 &  0.674(89)& 1.818(147)& 209\\
impnaphas&      N/A& 94$\%$& 15 &  0.663(69)& 1.370(114)& 281\\
\hline
\end{tabular}
\\$\mu=0.2$
\end{center}

In these tables ``has'' denotes Hasenbusch acceleration, ``imp'' denotes usage of the preconditioner and ``nap'' refers to the non area preserving update.


\end{document}